\documentstyle[dina4,epsf,12pt]{article}

\newcounter{saveeqn}%
\newenvironment{mathletters}%
{\setcounter{saveeqn}{\value{equation}}
\setcounter{equation}{0}%
\refstepcounter{saveeqn}%
}%
{\setcounter{equation}{\value{saveeqn}}}

\renewcommand{\date}[1]{}
\renewcommand{\maketitle}{}
\newcommand{\draft}{}
\newcommand{\pacs}[1]{}

\renewcommand{\title}[1]{\begin{center} {\LARGE #1} \\[2cm] \end{center}}
\renewcommand{\author}[1]{\begin{center} {\large #1} \\[1cm] \end{center}}
\newcommand{\address}[1]{\begin{center} {\large #1} \end{center} \vspace{4cm}}
\newcommand{\overdots}[1]{\stackrel{{\,\bf...}}{#1}}
\newcommand{\text}[1]{\rm #1}

\newcommand{\pdiff}[2]{\frac{\partial #1}{\partial #2}}

\newcommand{\rK}[1]{\left( #1 \right)} 
\newcommand{\eK}[1]{\left[ #1 \right]} 
\newcommand{\gK}[1]{\left\{ #1 \right\} } 

\begin{document}
\draft 
\title{Hierarchy of equations of motion for nonlinear coherent
excitations applied to magnetic vortices} 
\author{F.~G.~Mertens$^*$, H.~J.~Schnitzer$^*$, and A.~R.~Bishop}
\address{Theoretical Division, Los Alamos National Laboratory, MS
B262, Los Alamos, New Mexico 87545 \protect \\$^*$Physikalisches
Institut, Universit\"at Bayreuth, D-95440 Bayreuth, Germany}
\date{\today} \maketitle
\begin{abstract}
Starting from a travelling wave ansatz we show successively that the
shape of a nonlinear excitation generally depends also on the
$1^{\text{st}}$, $2^{\text{nd}}$, \ldots time derivative of the
position $\vec{X}$ of the excitation.  From the Hamilton equations we
derive a hierarchy of equations of motion for $\vec{X}$. The type of
the excitation determines on which levels the hierarchy can be
truncated consistently: "Gyrotropic" excitations are governed by
odd-order equations, non-gyrotropic ones by even-order
equations. Examples for the latter case are kinks in 1-dimensional
models and planar vortices of the 2-d anisotropic (easy-plane)
Heisenberg model.  The non-planar vortices of this model are the
simplest gyrotropic example. For this case we solve the Hamilton
equations for a finite system with one vortex and free boundary
conditions and calculate the parameters of the $3^{\text{rd}}$-order
equation of motion. This equation yields trajectories which are a
superposition of two cycloids with different frequencies, which is in
full agreement with computer simulations of the full many-spin model.
Finally we demonstrate that the additional effects from the
$5^{\text{th}}$-order equation are negligible.
\end{abstract}
 
 

\section{Introduction}
\label{sec:intro}
Nonlinear coherent excitations, such as solitons or solitary waves,
usually have some particle-like properties.  E.g., the equation of
motion of their `center-of-mass' is Newtonian, at least in a first
approximation. For the case of zero force the excitations can move at
constant velocity due to their inertial mass.  However, there are also
other excitations, which do not behave like classical particles, 
i.e., Galilei's law is not valid. For spin systems, good examples 
are certain magnetic domains and non-planar vortices in two- or 
three-dimensional models. For this type of nonlinear collective 
excitations Thiele
\cite{Thiele73,Thiele74} derived a $1^{\text{st}}$-order equation of
motion for the collective variable
$\vec{X}(t)$
\begin{equation}\label{eq:gyro} 
   \dot{\vec{X}} \times \vec{G} = \vec{F} \:,
\end{equation}
which we will refer to as the Thiele Eq.\ in the following. It is
valid only for steady-state motion because it was derived from the
Landau-Lifshitz equation for spins assuming a rigid shape of the
excitations.

$\vec{F}$ is either an external force or the force due to interactions
with other excitations. $\vec{V} \times \vec{G}$ is called a
gyrocoupling force. It is formally equivalent to the Lorentz
force. However $\vec{G}$ is not an external magnetic field but an
intrinsic quantity, produced by the excitation itself and carried
along with it. $\vec{G}$ is called the gyrocoupling vector, or for
short gyrovector. We call excitations with $|\vec{G}| = G \neq 0$
gyrotropic. For 1-d spin models $\vec{G}$ is always zero. For the
vortices of the 2-d anisotropic Heisenberg
model
\begin{equation}\label{eq:heis} 
   \vec{G} = 2\pi qp \vec{e}_{z}
\end{equation}
as was first calculated by Huber \cite{Huber82}. Here $\vec{e}_z$ is the
unit vector perpendicular to the $xy$-plane in which the magnetic ions
are situated. $q = \pm 1, \pm 2,\ldots$ is the vorticity, $p$ is a
second topological charge which is defined as the value of the spin
component $S^z$ at the vortex center in the continuum limit.

The 2-d anisotropic Heisenberg model with $XY$-symmetry is defined by
the spin Hamiltonian
\begin{equation}
\label{eq:hamiltonian}
   H = - J \sum_{<m,n>} \left[ S^x_m S^x_n + S^y_m S^y_n + (1-\delta)
      S_m^zS_n^z \right]
\end{equation}
with $0 < \delta \le 1$. Here ${<}m,n{>}$ labels the nearest-neighbor
sites of a square lattice. We treat the spin $\vec{S}$ as a classical
vector and set $S = J = 1$.

Due to the anisotropy $\delta$ the spins prefer to be oriented in the
$xy$-plane which is therefore often called easy plane. For strong
anisotropy $(\delta_c < \delta \le 1$, with $\delta_c \simeq 0.28)$
only planar vortices are stable \cite{Gouvea89,Wysin94}. In the static
case their spins are lying completely in the plane, while dynamically
small $S^z$-components develop. But at the vortex center $S^z$ is
always zero, thus $p = 0$ and $G = 0$ due to
(\ref{eq:heis}). Therefore the Thiele Eq.\ is not valid here.

In contrast to this case, for weak anisotropy $(0 < \delta <
\delta_c)$ only non-planar vortices are stable \cite{Gouvea89}. They
exhibit a localized structure of the $S^z$-components around the
vortex center, at which $S^z = \pm 1$. Although this structure changes
due to motion, the value at the center remains the same (section
\ref{sec:EQM2}). Thus $p = \pm 1$ determines to which side the
out-of-plane structure of the vortex is oriented and is therefore be
termed polarization. The Thiele Eq.\ is valid for steady-state motion
when the vortex shape is rigid (in the moving frame).  This includes
the case of a constant rotation on a circle.

In 1994 Wysin et al. \cite{Wysin_pl} dropped the rigid-shape
assumption by allowing the vortex shape to depend on the velocity
$\dot{\vec{X}}(t)$ and derived a generalized Thiele equation
\begin{equation}
\label{eq:wysin}
M\ddot{\vec{X}} + \dot{\vec{X}} \times \vec{G} = \vec{F}
\end{equation} 
As the velocity dependent parts of the vortex structure decay like
$1/r$ with the distance $r$ from the vortex center \cite{Gouvea89},
the mass was predicted to be
\begin{equation}
   \label{eq:goueva}
   M \simeq \frac{\pi}{4\delta} \ln(L/a_0)
\end{equation}
where $L$ and $a_0$ are upper and lower cut-offs in the order of the
system size and lattice constant, respectively. The same mass also
appears in the kinetic energy $M\dot{X}^2/2$ of a vortex
\cite{Gouvea89}.

According to Eq.\ (\ref{eq:wysin}) the trajectory $\vec{X}(t)$ of the
vortex center is formally the same as that of an electric charge $e$
in a plane with a perpendicular magnetic field $\vec{B}$ and an
in-plane electric force $\vec{F}$, namely a cycloid with frequency
$\omega = G/M$, cf. the cyclotron frequency $eB/(Mc)$, where $c$ is
the speed of light. However, a test of this prediction by computer
simulations turned out to be rather difficult: For the model
(\ref{eq:hamiltonian}) the use of an external force creates additional
nonlinear excitations; e.g., an in-plane magnetic field creates a
double domain wall which connects the vortex with a boundary
\cite{Gouvea90}. Therefore 2-vortex simulations were performed where
each vortex is driven by the force between them
\cite{Mertens94,Voelkel94}.  Large square systems with free boundaries
were used and the trajectories were chosen such that the vortices
stayed far away from each other and from the boundaries.

A 2-vortex theory was developed \cite{Voelkel94} which explains
one very important qualitative feature of the simulations: 
There are four main scenarios in vortex dynamics, however 
cycloidal oscillations around a mean trajectory $\vec{X}^0 (t)$ can 
be observed only for
two of them, namely where either a vortex and an antivortex (i.e.\ $q_1=-q_2$)
rotate around each other, or where two vortices with equal vorticity
perform a parallel translation with constant average speed.
There are no oscillations for the other 2 scenarios of
`vortex-vortex rotation' and `vortex-antivortex translation'.

However, the frequency $\omega$ of the observed cycloidal oscillations
yielded a mass\footnote{More precisely, for a 2-vortex system an
effective mass tensor was defined, its eigenvalues were both much
larger than predicted.} $M = G/\omega = 2 \pi/\omega$ which was much
larger than predicted by Eq.\ (\ref{eq:goueva}), see Table Ib of
Ref. \cite{Voelkel94}.  Later a second severe discrepancy was
discovered after having improved the simulations for the 2-vortex
rotation \cite{Mertens95}: Better initial conditions were used and
instead of a square a circular system was used, with symmetric initial
positions. Here for each vortex the mean trajectory $\vec{X}^o(t)$ is
a circle; the spectrum of the oscillations can be measured much more
accurately and shows very clearly two frequencies $\omega_{1, 2}$ with
about the same strength, instead of the one frequency $\omega = G/M$
predicted by Eq.\ (\ref{eq:wysin}).

Because of the above two discrepancies between theory and simulations
we now derive a new theory which amounts to a hierarchy of equations of
motion for a nonlinear coherent excitation in a system with an
arbitrary Hamiltonian (section \ref{sec:EQM}).  
The derivation is completely general, but for
simplicity we consider a Hamiltonian which is a functional of only one
field and its canonical momentum; this is the case of our model
(\ref{eq:hamiltonian}) in the continuum limit.

The level $n$ in this hierarchy is defined as the order $n$ of the
highest time derivative which appears in the equation of motion. A
classification of the excitations determines on which levels the
hierarchy can be truncated consistently: The dynamics of gyrotropic
excitations is described by odd-order equations, the simplest example
is the non-planar vortex of model
(\ref{eq:hamiltonian}). Non-gyrotropic excitations (e.g.  kinks in 1-d
models, planar vortices in model (\ref{eq:hamiltonian})) are governed
by even-order equations of motion.

In order to obtain the above classification one must know at least the
order of magnitude of the parameters in the equations of motion. For
the calculation of the parameters it is necessary to solve the
Hamilton equations for a system with one excitation, choosing
appropriate boundary conditions. We do this for our model 
(\ref{eq:hamiltonian}) in section \ref{sec:EQM2} using a finite system with 
free boundaries.

After the calculation of the parameters the equation of motion can be
solved. For the non-planar vortices we solve a $3^{\text{rd}}$-order
equation, which in fact yields the observed frequencies $\omega_{1,
2}$ (the trivial solution $\omega = 0$ yields the mean trajectory
$\vec{X}^o(t)$), section \ref{sec:mass}. 
We compare with new simulations where a {\em single}
vortex on a circle is driven 
by its image vortex (section \ref{sec:comp}).
We only note that the 2-vortex simulations yield a similar spectrum 
but cannot be compared quantitatively with this 1-vortex theory.
We also show that the $5^{\text{th}}$-order equation, which appears
on the next consistent level of the hierarchy, predicts two additional
frequencies which can also be observed, but they are very
weak. Therefore all levels higher than three can safely be neglected
in this case.
 
In section \ref{sec:disc} we discuss our results and in the appendix
we describe the numerical part of this work in more detail.
 
\section{Hierarchy of equations of motion}
\label{sec:EQM}
We consider an arbitrary classical Hamiltonian which is a functional
of a field $\phi (\vec{r}, t)$ and its canonical momentum $m(\vec{r},
t)$. We introduce a collective variable $\vec{X}(t)$ for the position
of a nonlinear coherent excitation. We make a travelling wave ansatz
$\phi(\vec{r}, t) = \phi(\vec{r} - \vec{X}(t))$, $m(\vec{r}, t) =
m(\vec{r} - \vec{X}(t))$ where, strictly speaking, the functions on
the r.h.s.\ should bear an index, which is omitted here for
simplicity.  We insert
\begin{equation}
   \label{eq26} \dot{\phi} = \pdiff{\phi}{X_j} \dot{X}_j\quad,\quad
   \dot{m} = \pdiff{m}{X_j} \dot{X}_j\quad,
\end{equation}
where a summation over $j = 1, 2, 3$ is implied, into the Hamilton
equations
\begin{equation}
\label{eq:delta}
   \dot{\phi} = \frac{\delta H}{\delta m} \quad,\quad \dot{m} =
   -\frac{\delta H}{\delta\phi} .
\end{equation}
Since the r.h.s. of these equations generally contain $m$ and $\phi$
it is clear that $m$ and $\phi$ depend not only on $\vec{X}$ but also
on $\dot{\vec{X}}$. Thus the shape of the collective excitations
generally depends on the velocity and we take this into account by the
improved ansatz $\phi(\vec{r}, t) = \phi(\vec{r} -
\vec{X},\dot{\vec{X}})$, $m(\vec{r}, t) = m(\vec{r} - \vec{X},
\dot{\vec{X}})$. However, if we now insert $\dot{\phi}$ and $\dot{m}$
into the eqs. (\ref{eq:delta}) we see an additional dependence on
$\ddot{\vec{X}}$, and so on. Obviously we must truncate somewhere. We
will see below that a truncation at the derivative $\vec{X}^{(n-1)}$
yields an $n^{th}$-order equation of motion. For gyrotropic
excitations only odd-order equations will turn out to be consistent
(section \ref{sec:mass}).  As an example we derive here the
$3^{\text{rd}}$-order equation by the ansatz
\begin{mathletters}
\label{ansatz:phi-m}
\begin{eqnarray}
   \label{eq28}
   \phi(\vec{r},t) &=& \phi(\vec{r} - \vec{X}(t), \dot{\vec{X}}(t),
   \ddot{\vec{X}}(t))\\
   \label{eq29}
   m(\vec{r},t) &=& m(\vec{r}-\vec{X}(t), \dot{\vec{X}}(t),
\ddot{\vec{X}}(t)).
\end{eqnarray}
\end{mathletters}%

It is unusual that a collective excitation should depend on the
acceleration, but it will be seen later that this dependence yields
one of the dominant terms in the equation of motion. This equation is
now easily derived by using the technique of Wysin et
al. \cite{Wysin_pl}. Our ansatz is inserted into the Hamilton
equations (\ref{eq:delta}) and yields
\begin{mathletters}
\begin{eqnarray}
\label{eq210}
   \pdiff{\phi}{X_j}\dot X_\j + \pdiff{\phi}{\dot X_j}\ddot X_j +
     \pdiff{\phi}{\ddot X_j}\overdots X_j &=& \phantom-\frac{\delta
     H}{\delta m}\\
\label{eq211}
   \pdiff{m}{X_j}\dot X_\j + \pdiff{m}{\dot X_j}\ddot X_j +
     \pdiff{m}{\ddot X_j}\overdots X_j &=& -\frac{\delta
     H}{\delta\phi},
\end{eqnarray}
\end{mathletters}%
where $H$ depends on $\vec{X}$, $\dot{\vec{X}}$ and $\ddot{\vec{X}}$
via $\phi$ and $m$. Multiplying these equations by $\pdiff{m}{X_i}$
and $\pdiff{\phi}{X_i}$, resp., subtracting the equations from each
other, and integrating over $\vec{r}$ we obtain
\begin{equation}
\label{eq212}
   {\bf A}\overdots{\vec X} + {\bf M} \ddot{\vec X} + {\bf G}\dot{\vec
  X} = \vec F\end{equation} Here
\begin{equation}
\label{eq213}
   F_i = -\int\!d^3r\, \pdiff{{\cal H}}{X_i}
\end{equation}
with the Hamiltonian density $\cal H$ is the static force, it is
either an external force or the force due to the interactions with
other nonlinear excitations.
\begin{equation}
\label{eq214}
   G_{ij} = \int\!d^3r\,\left\{\pdiff{\phi}{X_i}\pdiff{m}{X_j} -
      \pdiff{\phi}{X_j}\pdiff{m}{X_i} \right\}
\end{equation}
is equivalent to the gyrocoupling tensor of Thiele
\cite{Thiele73,Thiele74}.  Because of the antisymmetry of $\bf G$ the
term $\bf G\dot{\vec{X}}$ can be written as
$-\vec{G}\times\dot{\vec{X}}$. The gyrovector $\vec{G}$ is orthogonal
to the plane defined by $\vec{X}$ and $\dot{\vec{X}}$ and has already
been discussed in the introduction. The terms ${\bf G}\dot{\vec{X}}$
and $\vec{F}$ constitute the Thiele Eq.\ 
valid only for steady-state motion. We note that the above method is
much shorter than the original one by Thiele, who started directly
from the Landau-Lifshitz Eq.\ (\ref{LLE}).

\begin{equation}
\label{eq215}
   M_{ij} = \int\!d^3r\,\left\{\pdiff{\phi}{X_i}\pdiff{m}{\dot X_j} -
      \pdiff{\phi}{\dot X_j}\pdiff{m}{X_i} \right\}
\end{equation}
is referred to as the mass tensor, and
\begin{equation}
\label{eq216}
   A_{ij} = \int\!d^3r\,\left\{\pdiff{\phi}{X_i}\pdiff{m}{\ddot X_j} -
      \pdiff{\phi}{\ddot X_j}\pdiff{m}{X_i} \right\}
\end{equation}
as the $3^{\text{rd}}$-order gyrotensor. In order to see which
properties they have it is necessary to calculate the functions $\phi$
and $m$ in the ansatz (\ref{ansatz:phi-m}) which describe the shape of
the excitations. We do this for our model (\ref{eq:hamiltonian}).

\section{Hamilton equations for a finite system with one vortex and free 
   boundaries}
\label{sec:EQM2}
We consider the 2-d anisotropic Heisenberg model
(\ref{eq:hamiltonian}) and introduce
\begin{mathletters}
\label{motion}
\begin{eqnarray}
   \phi(\vec{r}, t) &=& \tan^{-1}\frac{S^y(\vec{r}, t)}{S^x(\vec{r},
   t)}\\ m(\vec{r}, t) &=& S^z(\vec{r}, t),
\end{eqnarray}
\end{mathletters}%
where the site index at the spin components has been replaced by the
variable $\vec{r} = (x_1, x_2)$.  The Hamiltonian is \cite{Gouvea89}
\begin{equation}
\label{vd1:HM:H2}
   H = \frac{1}{2} \,\int d^2r\gK{(1-m^2)(\nabla\phi)^2 +
\delta[4m^2-(\nabla m)^2] + \frac{(\nabla m)^2}{1-m^2}}
\end{equation}
and the Hamilton equations read
\begin{mathletters}
\begin{eqnarray}
\label{eq320}
   \frac{\partial\phi}{\partial t} & = & m[4\delta-(\nabla\phi)^2]
         -\rK{\frac{1}{1-m^2}-\delta} \Delta m - \frac{m(\nabla
         m)^2}{(1-m^2)^2} \\%
\label{eq321}
   \frac{\partial m}{\partial t} & = & (1-m^2)\Delta\phi-2m\nabla
    m\nabla\phi.
\end{eqnarray}
\end{mathletters}%

For $0<\delta<\delta_c$ (with $\delta_c\approx0.28$ for a square
lattice) only non-planar vortices are stable and they have the
following static structure \cite{Gouvea89}
\begin{eqnarray}
\label{eq322}
   \phi_0 &=& q\tan^{-1}\frac{x_2'}{x_1'} + \varphi_0 \\
\label{eq323}
   m_0 &=& p a_2\sqrt{\frac{r_V}{r'}} e^{-r'/r_V}\quad\mbox{for $r'\gg
r_V$} \\
\label{eq324}
   m_0 &=& p\eK{1-a_1^2\rK{\frac{r'}{r_V}}^2}\quad\mbox{for $r'\ll
   r_V$}
\end{eqnarray}
with
\begin{equation}
\label{eq325}
   \vec r\,' = \vec r - \vec X.
\end{equation}
The polarization is $p=\pm1$ and only vorticities $q=\pm1$ are
considered. The constants $a_1$ and $a_2$ can be determined by
matching the two solutions (\ref{eq323}) and (\ref{eq324}) at some
intermediate distance, e. g. at $r\approx r_V$. The characteristic
length
\begin{equation}
\label{eq326}
   r_V = \frac{1}{2} \sqrt{\frac{1-\delta}{\delta}}
\end{equation}
is interpreted as the radius of the vortex core.  
The constant $\varphi_0$ in (\ref{eq322}) stays arbitrary.
From the static structure one obtains for (\ref{eq214})
\begin{equation}
\label{eq327}
    G_{ij} = G \, \epsilon_{ij}
\end{equation}
with $G=2\pi q p$. $\epsilon_{ij}$ is the 2-d completely antisymmetric
tensor.

For a slowly moving vortex we expand in a perturbation series
\begin{mathletters}
\begin{eqnarray}
\label{eq328}
   \phi &=& \phi_0(\vec r\,') + \phi_1(\vec r\,',\dot{\vec X}) +
               \phi_2(\vec r\,',\ddot{\vec X}) \\
\label{eq329}
   m &=& m_0(\vec r\,') + m_1(\vec r\,',\dot{\vec X}) + m_2(\vec
               r\,',\ddot{\vec X})
\end{eqnarray}
\end{mathletters}%
and assume that the first and second order terms depend linearly on
$\dot X_j$ and $\ddot X_j$, resp.  Slow motion means here that the
velocity is much smaller than the spin-wave velocity which is
$2\sqrt{\delta}$ in our dimensionless units \cite{Gouvea90}.

In the next section it will be seen that the main contributions to the
integrals for ${\bf M}$ and ${\bf A}$ stem from the region far away
from the vortex center, i.e. from $r'\gg r_V$.  Particularly the
dependence on the system size naturally comes entirely from this
region. Therefore it is sufficient to solve the Hamilton equations
only for this region. We assume that here only the terms $4\delta m$
and $\Delta\phi$ on the r.h.s. of (\ref{eq320}) and (\ref{eq321}),
resp., are important for the dynamic parts of $\phi$ and $m$. This
assumption can be justified a posteriori.

In $O(\dot X_j)$ we then obtain for $r'\gg r_V$
\begin{equation}
\label{eq330}
   \pdiff{\phi_0}{X_j}\dot X_j = 4\delta m_1 \quad,\quad
   \pdiff{m_0}{X_j}\dot X_j = \Delta\phi_1 .
\end{equation}
The first equation gives
\begin{equation}
\label{eq331}
   m_1 = \frac{q}{4\delta(r')^2} \rK{{x_2'}\dot{X}_1 -
         {x_1'}\dot{X}_2} = \frac{q\sin{\varphi'}}{4\delta
         r'}|\dot{\vec{X}}|,
\end{equation}
where $\varphi'$ is the angle between $\vec{r}\,'$ and
$\dot{\vec{X}}$.  For the second equation in (\ref{eq330}) the
inhomogenity vanishes at large distances. We therefore have the
solution
\begin{equation}\label{eq332}
   \phi_1 = p c x_j' \dot X_j
\end{equation}
for large $r'$ which increases linearly towards the free boundaries,
in contrast to the asymptotic solution of ref.\ \cite{Gouvea89} where
an infinite system with decaying boundary conditions was
considered. The constant $c$ can be determined by matching the
solution (\ref{eq332}) to the small-$r'$ solution of ref.\
\cite{Gouvea89}.

In $O(\ddot X_j)$ we obtain by the above methods
\begin{equation}
\label{eq333}
   m_2 = p b x_j'\ddot X_j
\end{equation}
with $b=c/(4\delta)$, and
\begin{equation}
\label{eq334}
   \phi_2 = \frac{q}{8\delta} \ln r' \rK{x_2'\ddot X_1-x_1'\ddot X_2}.
\end{equation}
Here $m_2 \sim\cos \chi'$ and $\phi_2 \sim \sin\chi'$, where $\chi'$
is the angle between $\vec r\,'$ and $\ddot{\vec X}$.

\begin{figure}
\centerline{\epsfxsize=8.0truecm \epsffile{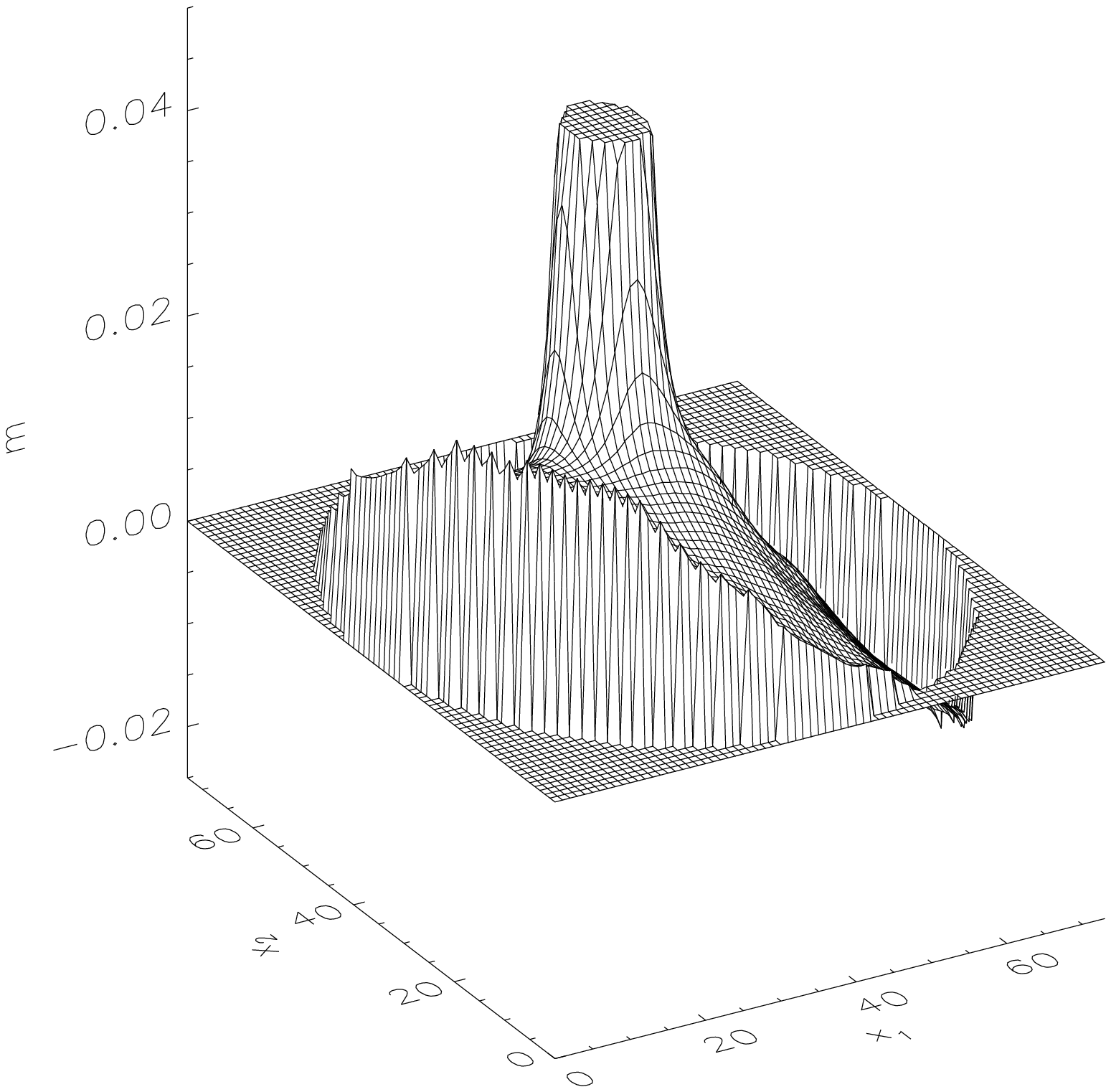}}
\caption{Out-of-plane structure of the vortex at the $7^{\text{th}}$
   turning point of the trajectory in Fig.\ \protect\ref{fig4}. Here
   the acceleration has a maximum and points in radial direction,
   while the velocity is small and points in azimuthal direction.}
\label{fig1}
\end{figure}

\begin{figure}
\centerline{\epsfxsize=8.0truecm \epsffile{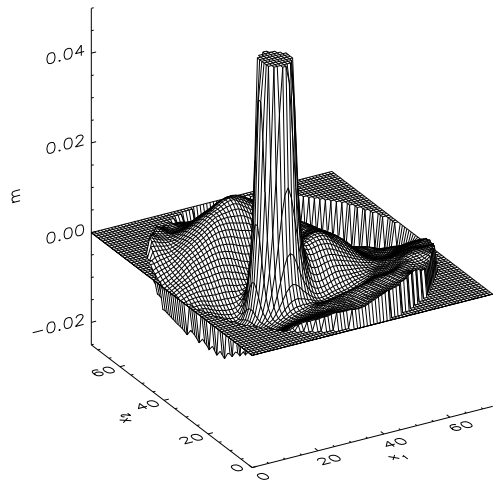}}
\caption{Out-of-plane structure at the middle between the
   $7^{\text{th}}$ and the $8^{\text{th}}$ turning point of Fig.\
   \protect\ref{fig4}.  Here the velocity has a maximun and points in
   radial direction, while the acceleration has a minimum.}
\label{fig2}
\end{figure}

\begin{figure}
\centerline{\epsfxsize=8.0truecm \epsffile{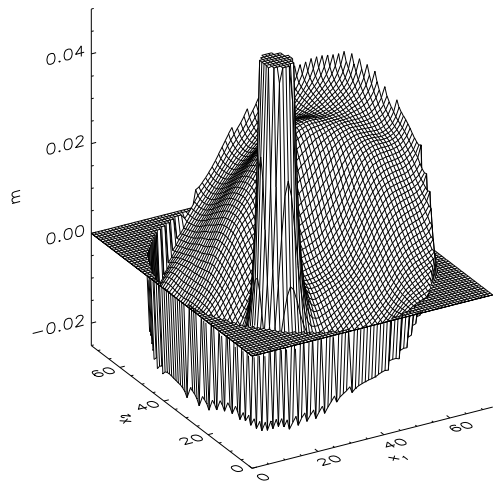}}
\caption{Same as in Fig.\ \protect\ref{fig1}, but at the
   $8^{\text{th}}$ turning point, where the accelaration points in
   negativ radial direction.}
\label{fig3}
\end{figure}

The results of this section can be tested by looking at snapshots of
the orientation of the spins in our computer simulations for the
discrete system (the simulations are discussed in section
\ref{sec:comp}). The dynamic parts of $\phi$ are very difficult to
observe because $\phi_0$ must be substracted first. But this depends
very sensitively on the vortex position, which is known only within a
certain accuracy. However, $m_0$ vanishes exponentially for $r'\gg
r_V$. Here the dynamic parts of $m$ can be observed directly. Though
$m_2$ increases linearly with $r'$ while $m_1$ decays, $m_2$ generally
cannot be distinguished clearly from $m_1$ because generally
$|\ddot{\vec X}|\ll|\dot{\vec X}|$.  In order to observe $m_2$
nevertheless, we have selected specific points of the vortex
trajectories in section \ref{sec:comp}: At the turning points the
acceleration has a maximum while the velocity is small. Here the
predicted $\cos\chi'$-dependence of $m_2$ and its linear increase
towards the boundaries are seen very clearly (Figs.~\ref{fig1} and
\ref{fig3}). In the middle between two turning points the trajectory
is nearly straight and the acceleration is small. Here $m_2$ is in
fact barely visible (Fig.~\ref{fig2}), instead several humps can be
observed which are probably produced by $m_1$.  By the evaluation of
contour plots of the above snapshots we estimate $b\approx2.5$ for the
parameter in (\ref{eq333}). This parameter is the only one we need for
the calculation of the integrals in the following section.

\section{Mass, third-order gyrotensor and solution of the equation of motion}
\label{sec:mass}
Since we consider very slow motion we need to calculate only the rest
mass, i.e. in Eq.\ (\ref{eq215}) the derivations with respect to $X_i$
are applied only to the static parts of $\phi$ and $m$. As $\phi_1$
and $m_1$ depend linearly on the velocity,
\begin{equation}
\label{eq435}
   M_{ij} = \int\!d^2r\,\left\{\pdiff{\phi_0}{X_i}\pdiff{m_1}{\dot
      X_j} - \pdiff{\phi_1}{\dot X_j}\pdiff{m_0}{X_i} \right\}
\end{equation}
is the rest mass. In the same way we obtain the constant part of
(\ref{eq216}) by
\begin{equation}
\label{eq436}
   A_{ij} = \int\!d^2r\,\left\{\pdiff{\phi_0}{X_i}\pdiff{m_2}{\ddot
      X_j} - \pdiff{\phi_2}{\ddot X_j}\pdiff{m_0}{X_i} \right\}.
\end{equation}
An analytic calculation is possible if we choose a circular system
(radius $L$) and consider a vortex with its center $\vec X$ at the
circle center, which is chosen as origin. We divide the integration
region into an inner part $0\le r\le a_c$ and an outer part $a_c\le
r\le L$, where we choose $a_c\gg r_V$.  The inner regions yield
$L$-independent contributions, while the contributions from the outer
region turn out to increase with $L$ and thus dominate for large
$L$. As $m_0$ decays exponentially according to (\ref{eq323}), the
second terms in (\ref{eq435}) and (\ref{eq436}) give no contribution
for the outer region. Therefore we need to calculate only the first
terms and obtain with (\ref{eq322}), (\ref{eq331}) and (\ref{eq333})
\begin{mathletters}
\begin{eqnarray}
\label{eq437}
   M_{ij} = M \delta_{ij} &,& M = \frac{\pi}{4\delta}\ln\frac{L}{a_c}
      + const. \\
\label{eq438}  
   A_{ij} = A \epsilon_{ij} &,& A = \frac{Gb}{4}\rK{L^2-a_c^2} +
      const.
\end{eqnarray}
\end{mathletters}%
where $\delta_{ij}$ is the 2-d unit matrix.  The constants are the
above $L$-independent contributions from the inner region. They depend
on $a_c$, but together with the other $a_c$-dependent terms they must
add up to $a_c$-independent constants $M_0$ and $A_0$ because the
final result
\begin{equation}
\label{MA}
   M = \frac{\pi}{4\delta}\ln L + M_0 \quad,\quad A = \frac{Gb}{4}L^2
   + A_0
\end{equation}
must not depend on how the integration region is divided.

If the vortex center is not at the circle center the situation is less
symmetric, but ${\bf M}$ is still diagonal (in a system with axes
parallel and orthogonal to $\vec X$) and the $A_{ii}$ are still
zero. The non-vanishing matrix elements must be calculated by
numerical integration.  However, in our simulations the mean vortex
trajectory is a circle with radius $R_0 \ll L$ around the circle
center. Therefore the differences between $M_{11}$ and $M_{22}$ and
those between $A_{12}$ and $-A_{21}$ are small and will be neglected
in the following.

For the solution of the equation of motion (\ref{eq212}) we consider a
small displacement from a mean trajectory $\vec X^0(t)$
\begin{equation}
\label{eq439}
   \vec X(t) = \vec X^0(t) + \vec x(t).
\end{equation}
We consider a situation where the force is always pointing in the
$X_1$-direction and expand to $1^{\text{st}}$ order around $X_1^0=R_0$
\begin{equation}
   F = F_0 + F_0'x_1.
\end{equation}
The mean trajectory is a parallel to the $X_2$-axis, namely
$X_1^0=R_0$, $X_2^0=V_0t$, where $V_0=F_0/G$ is positive for a vortex
with $qp=1$.

For the displacement $\vec x$ we get two linear equations
\begin{mathletters}
\begin{eqnarray}
\label{eq441}
   A \overdots x_2 + M \ddot x_1 + G \dot x_2 &=&F_0' x_1 \\
\label{eq442}
  -A \overdots x_1 + M \ddot x_2 - G \dot x_1 &=& 0
\end{eqnarray}
\end{mathletters}%
which are solved by
\begin{mathletters}
\begin{eqnarray}
\label{eq443}
   x_1 &=& a_1\cos\omega_1 t + a_2\cos\omega_2 t \\
\label{eq444}
   x_2 &=& b_1\sin\omega_1 t + b_2\sin\omega_2 t
\end{eqnarray}
\end{mathletters}%
with
\begin{equation}
\label{eq445}
   \omega_{1, 2}^2 = \frac{2GA+M^2}{2A^2} \mp
     \sqrt{\frac{(4GA+M^2)M^2}{4A^4} + \frac{MF_0'}{A^2}}.
\end{equation}
The amplitude ratios $\kappa_{\alpha}=b_{\alpha}/a_{\alpha}$ are
obtained from
\begin{mathletters}
\begin{eqnarray}
\label{eq446}    
 (G - A \omega_{\alpha}^2)\, \omega_{\alpha}\kappa_{\alpha} &=& M
\omega_{\alpha}^2 + F_0'\\
\label{eq447}
G - A \omega_{\alpha}^2 &=& M \omega_{\alpha}\kappa_{\alpha}
\end{eqnarray}
\end{mathletters}
which yields
\begin{equation}
\label{eq448}
\kappa_{\alpha} = \pm \sqrt{1 + F_0^{\prime}/(M
\omega_{\alpha}^2)},\quad \alpha = 1, 2.
\end{equation}
In order to facilitate the discussion we set $F_0'=0$ (our simulations
are made anyway for the case of small $F_0$ and $F_0'$. Then we get
\begin{equation}
\label{eq449}
   \omega_{1, 2} = \sqrt{\frac{G}{A} + \rK{\frac{M}{2A}}^2} \mp
           \frac{M}{2A}
\end{equation}
with $\kappa_{1, 2}=\pm1$.  As $A\sim L^2$ while $M\sim \ln L$ for a
large system, the leading $L$-dependence of $\omega_{\alpha}$ is
$1/L$.

The frequencies $\omega_{1, 2}$ form a narrow doublet. The mean
frequency depends on $A$, but not on $M$
\begin{equation}
\label{eq450}
   \omega_c = \sqrt{\omega_1\omega_2} = \sqrt{G/A} \sim 1/L.
\end{equation}
Contrary to this, the splitting of the doublet
\begin{equation}
\label{eq451}
   \Delta\omega = \omega_2 - \omega_1 = M/A \sim \frac{\ln L}{L^2}
\end{equation}
is proportional to the mass.

We now discuss the size dependence of the different terms in the
equation of motion (\ref{eq212}). As every time derivative contributes
a factor $\omega_{\alpha} \sim 1/L$, we see that $A\overdots{X}_i \sim
1/L$ and $M\ddot{X}_i \sim \ln L/L^2$ for a large
system. There\-fore $A\overdots{X}_i$ cannot be neglected when
$M\ddot{X}_i$ is retained. This is the reason for the discrepancies
resulting from the $2^{\text{nd}}$-order equation (\ref{eq:wysin}),
see the introduction. However, the neglection of both
$A{\overdots{X}}_i$ and $M\ddot{X}_i$ represents a consistent
approximation, namely the Thiele Eq.\ (\ref{eq:gyro}).

We have just seen that the levels $n = 1$ and $n = 3$ represent
consistent approximations in a hierarchy of equations of motion. Let
us now include two additional time derivatives in our ansatz
(\ref{ansatz:phi-m}) which yields two additional orders in the
equation of motion. Using the methods of sections \ref{sec:EQM2} and
\ref{sec:mass} one can show that the parameters in the
$5^{\text{th}}$- and $4^{\text{th}}$-order terms scale like $L^4$ and
$L^2\ln L$, resp.. The same arguments as above show that $n = 5$
is the next consistent level. Here we get 4 frequencies
$\omega_{\alpha}$ (besides the trivial solution $\omega =
0$). Neglecting for the moment the small $2^{\text{nd}}$- and
$4^{\text{th}}$-order terms, we obtain two two-fold degenerate
frequencies. The inclusion of the neglected terms lifts the degeneracy
and gives a spectrum of two doublets $\omega_{1, 2}$ and $\omega_{3,
4}$. The corresponding amplitudes $a_{\alpha}$ are free constants in a
general solution which are determined by the initial conditions (only
the ratios $\kappa_{\alpha} = b_{\alpha}/a_{\alpha}$ are fixed, as in
(\ref{eq448})). It will turn out in the next section that $a_{3, 4}$
are generally so small that $\omega_{3, 4}$ cannot be observed.  Only
for very special conditions can $\omega_{3, 4}$ be seen.

Finally we shortly discuss non-gyrotropic excitations which we define
as having only even-order terms in the equations of motion. Examples
are the kinks in the 1-d nonlinear Klein-Gordon models for which $\bf
{G}$ and $\bf {A}$ vanish. Here a $4^{\text{th}}$-order equation was
derived but not considered in detail \cite{Boesch88}. Another example
are the planar vortices of the 2-d anisotropic Heisenberg model
(stable for $\delta>\delta_c$).  Here $G = 0$ in Eq.\
(\ref{eq:heis}) because of $p = 0$; all other odd-order terms also
vanish because they are proportional to $p$, e.g. (\ref{eq436}).

\section{Comparison with simulations}
\label{sec:comp}
A single vortex with polar coordinates $(R,\phi)$ on a circle with
radius $L$ has an image vortex at $R_i = L^2/R$ and the same $\phi$
(as in 2-d electrostatics). For free boundaries the image has opposite
vorticity but the same polarization \cite{Mertens94} and the force is
$F(R) = 2\pi/(R_i-R)$. The equation of motion (\ref{eq212}) has a
steady-state solution: a constant rotation on a circle $R = R_0,\,
\phi(t) = \omega_0t$ with the following relation between $R_0$ and
$\omega_0$:
\begin{equation}
\label{eq552}
-A\omega_0^3 R_0 - M\omega_0^2R_0 + G\omega_0R_0 = F(R_0).
\end{equation}
This circle is identified with the mean trajectory around which we
observe oscillations in the simulations (details are given in the
Appendix). In Fig.~\ref{fig4} the radial coordinate $R(t)$ is plotted
vs.\ the coordinate $R_0\phi(t)$ in azimuthal direction. This figure
already shows qualitatively that two frequencies are involved. In
fact, the Fourier spectra for $r(t) = R(t) - R_0$ and $\varphi(t) =
\phi(t) -\omega_0t$ in Fig.~\ref{fig5} clearly show two dominant
frequencies $\omega_{1, 2}$. The phase shifts $\delta_{1, 2}$ are
approximately $\pi/2$ and $-\pi/2$ (Table \ref{table1}).  Thus the
simulation data are described very accurately by
\begin{mathletters}
\label{eq:r_phi}
\begin{eqnarray}
   \label{eq553}
   r(t) &=& a_1\, \mbox{cos}\, \omega_1t + a_2\, \mbox{cos}\,
   \omega_2t\\
   \label{eq554}
   R_0 \varphi(t) &=& b_1\, \mbox{sin}\, \omega_1t + b_2\,
   \mbox{sin}\, \omega_2t
\end{eqnarray}
\end{mathletters}%
with
\begin{equation}
\label{eq555}
\mbox{sign}\, \kappa_{\alpha} = \pm 1 \quad, \alpha = 1, 2.
\end{equation}
As there are displacements not only in radial direction but also in
azimuthal direction the trajectories are a superposition of two
cycloids with frequencies $\omega_{1, 2}$.

In order to compare with our theory we must solve the equation of
motion (\ref{eq212}) in polar coordinates. The result agrees
completely with (\ref{eq:r_phi}), (\ref{eq555}). The formula for
$\omega_{\alpha}$ is much more complicated than the cartesian result
(\ref{eq445}) due to many additional, $\omega_0$-dependent
terms. However, $\kappa_{\alpha}$ can be written in a rather simple
form by using (\ref{eq552})
\begin{eqnarray}
\label{eq556}
   \kappa_{\alpha}^2 &=& 1 + \frac{f}{(M + 3 A \omega_0)
   \omega_{\alpha}^2}\\
\label{eq557}
   f &=& F_0' - \frac{F_0}{R_0} =
   \frac{2\pi}{L^2}\frac{2\eta^2}{(1-\eta^2)^2}
\end{eqnarray}
with $\eta = R_0/L$. For $\eta \ll 1$, which is the case in our
simulations, a short calculation shows that $\kappa_\alpha=\pm 1$,
neglecting terms of order $\eta^2$ and higher. In this case it turns
out that the substitution
\begin{equation}
\label{eq558}
   \omega_{\alpha} \pm \omega_0 \rightarrow \nu_{\alpha} \quad, \alpha
   = 1, 2
\end{equation}
transforms the complicated eigenvalue equations for $\omega_{\alpha}$
into simpler ones
\begin{mathletters}
\begin{eqnarray}
   \label{eq559} (G - A \nu_{\alpha}^2) \nu_{\alpha} \kappa_{\alpha}
   &=& M \nu_{\alpha}^2 + F_0'\\ \label{eq560} (G - A \nu_{\alpha}^2)
   \nu_{\alpha} &=& (M\nu_{\alpha}^2 + F_0/R_0) \kappa_{\alpha}.
\end{eqnarray}
\end{mathletters}%
Here the difference between $F_0'$ and $F_0/R_0$ must be neglected to
be consistent with the above neglection of $f$ in (\ref{eq556}).  We
note in passing that this difference and the one between $|\kappa_1|$
and $|\kappa_2|$ are responsible for the deviations of ${\bf M}$ and
${\bf A}$ from the symmetries in (\ref{eq437}) and (\ref{eq438}),
which are discussed below these equations.

As we know only the size dependence of $M$ and $A$, it does not make
sense to solve (\ref{eq559}), (\ref{eq560}) for
$\nu_{\alpha}$. Instead, we calculate $M$ and $A$ as functions of
$\nu_{\alpha}$, or $\omega_\alpha$ and $\omega_0$, which have been
measured in simulations for different system sizes:
\begin{mathletters}
\begin{eqnarray}
   \label{eq561} M &=& (B_1 \nu_2^3 - B_2 \nu_1^3)/D\\ \label{eq562} A
   &=& (B_2\kappa_1 \nu_1^2 - B_1 \kappa_2\nu_2^2)/D
\end{eqnarray}
\end{mathletters}%
with
\begin{mathletters}
\begin{eqnarray}
   \label{eq563} B_\alpha &=& G\nu_\alpha - \kappa_\alpha F_0/R_0\\
   \label{eq564} D &=& \nu_1^2 \nu_2^2 (\kappa_1 \nu_2 - \kappa_2
   \nu_1).
\end{eqnarray}
\end{mathletters}%
The results in Table \ref{table2} have to be compared with (\ref{MA}),
which was calculated for $R_0=0$, however.  Therefore we take only the
data for small $R_0/L=\eta$, e.g.\ those for $\eta\approx0.17$. The
data for $A$ are well represented by $A=A_0+CL^\alpha$ with
$\alpha=2.002$, $C=4.67$ and $A_0=40$. This agrees perfectly with the
$L^2$-dependence in (\ref{MA}). From $C$ we obtain $b=2.97$, which
agrees rather well with the value 2.5 which was estimated at the end
of section 3. However, the values for $M$ are practically independent
of $L$, in contrast to the logarithmic dependence in (\ref{MA}). This
can be explained by the following: In contrast to the $1/r'$-decay in
(\ref{eq331}), $m_1$ seems to go to an $L$-dependent constant at the
boundary (however, this is difficult to see because only $m_1+m_2$ is
observed, and $m_2$ is never exactly zero).  At this point we must
realize that Eq.\ (\ref{eq331}) is only an approximation, because for
$\phi_0$ in (\ref{eq330}) we have taken (\ref{eq322}), i.e., the
contribution of the image vortex has been neglected. This effect and
others, such as the splitting of the $M_{ii}$-components for $R_0\ne0$
and the influence of a second vortex in the system, will be considered
in forthcoming work.

The last point in this paper is the observation of additional
frequency doublets which can be related to higher levels of the
hierarchy (see the penultimate paragraph of section 4).  We have
performed simulations for a vortex in the center of a square system
with antiperiodic boundary conditions.  In this particular case there
are many image vortices at positions which can be calculated as in
electrostatics. The forces from all the images cancel exactly,
therefore the vortex in the center is influenced only by the small
pinning forces from the lattice.  We observe that the vortex center
moves on small-amplitude cycloidal trajectories, which are either
completely inside a lattice cell or which go over severall cells.
However, the spectrum is always practically the same which means that
the pinning forces do not substantially influence the frequencies.  In
Fig.~\ref{fig6} we clearly see not only a second doublet but also a
third one, which we want to relate to the $5^{\text{th}}$ and
$7^{\text{th}}$ levels of the hierarchy, resp. In order to show this
relation we must check that the parameters $M$ and $A$ do not change
dramatically when we go from the $3^{\text{rd}}$ to the
$5^{\text{th}}$ level, for instance. On this level we get a
$4^{\text{th}}$-order eigenvalue equation for $\omega_\alpha$, after
splitting off the trivial solution. We need not solve this equation
because we want to calculate $M$, $A$, $\ldots$ as functions of the
$\omega_\alpha$'s.  This is achieved by using Vieta's rules which
express the coefficients of an $n^{\text{th}}$-order polynominal
equation by its roots.  Table \ref{table3} contains the observed
$\omega_\alpha$'s and the resulting values for $M$ and $A$, which in
fact do not differ much on the different levels of the hierarchy. We
remark that this test is very sensitive: Small changes in
$\omega_\alpha$ result in large changes in the values of $M$ and $A$.

Finally we would like to stress that the additional $\omega_\alpha$'s
can usually not be observed in the spectra because their amplitudes
are too weak, thus our $3^{\text{rd}}$-order equation of motion is
normally sufficient to describe the simulations.

\section{Discussion}
\label{sec:disc}
Inserting the usual travelling-wave ansatz into the Hamilton equations
shows that the shape of a nonlinear coherent excitation generally
depends also on its velocity $\dot{\vec X}$. Iterating this process we
have shown that there is an additional dependence on $\ddot{\vec X}$,
$\overdots{\vec X}$, and so on. This yields a hierarchy of equations
of motion for $\vec X(t)$. For the case of the non-planar vortices of
the 2-d anisotropic Heisenberg model the odd levels of this hierarchy
represent an increasingly better description of the dynamics as
observed in computer simulations of the full spin system.

Naturally the question arises why spin waves do not appear here,
neither in theory nor in simulations. As to the latter point, we make
two remarks: 1.) For a short time after the start of a simulation spin
waves are radiated because the initial condition usually does not
perfectly represent a moving vortex. In fact, if we use approximate
formulas for the vortex structure, the generated spin waves are
clearly observed. In our older simulations \cite{Mertens94,Voelkel94} 
we eliminated them 
by adding a Gilbert damping term to the Landau-Lifshitz Eq.\ for the
above short time period. However, in the present paper we use an
iterative method (see Appendix) which produces a stationary vortex
solution which is so good that practically no spin waves are
generated.  2.) One might think that a vortex would continously
radiate spin waves because it is subject to accelerations on its
complicated cycloidal trajectory. Even if the amplitudes of these spin
waves were too small to observe them directly, the effect should be
detected indirectly by an energy loss of the vortex. Interestingly, in
our simulations, even after a long time (several periods
$T_0=2\pi/\omega_0$, see (\ref{eq552})) there was no detectable energy
loss; this suggests the possibility of additional conservation laws.

As to the theory, there is an alternative formalism \cite{Schnitzer_thesis} 
which includes spin waves. 
Instead of eqs.\ (\ref{eq28}), (\ref{eq29}) the following ansatz is used
\begin{mathletters}
\label{eq:phi_m}
\begin{eqnarray}
   \phi(\vec r,t)&=&\phi(\vec r-\vec Y(t)) + \xi(\vec r,t) \\ m(\vec
      r,t)&=& m(\vec r-\vec Y(t)) + \chi(\vec r,t)
\end{eqnarray}
\end{mathletters}%
where the functions $\phi$ and $m$ on the r.h.s.\ represent the static
structure of a vortex, while $\xi$ and $\chi$ represent spin waves. As
the number of degrees of freedom on the r.h.s.\ of eqs.\ (\ref{eq:phi_m}) is
larger than on the l.h.s., constraints must be introduced which have
the form of orthogonality relations \cite{Tomboulis75,Boesch88}.

One of the constraints leads to an implicite definition of the vortex
position $\vec Y$. This definition is global in the sense that
all spins in the system are
involved, in contrast to the local definition of the vortex center
$\vec X$ that has been used in the present paper (see
Appendix). Consequently different trajectories $\vec Y(t)$ and
$\vec X(t)$ are obtained by analyzing the same simulation data. $\vec
Y(t)$ turns out to be equivalent to the mean trajectory $\vec R_0(t)$
belonging to the cycloidal trajectories $\vec X(t)$ obtained in
the present paper.

In the alternative formalism the cycloidal trajectories are
obtained by a coupling of $\vec Y(t)$ to certain quasi-local spin wave
modes.  Far away from the vortex center these modes are extended while
they are localized in the core region \cite{Wysin95}. 
Quasi-local eigenmodes are in contrast to the truly localized,
intrinsic modes of e.g.\ the kinks in the $\phi^4$-model
\cite{Bishop80}.
An exact numerical
diagonalization for a small system ($L=20$) with one vortex shows that
essentially only two modes are involved. Their frequencies are
identical to those of the doublet which we observe in the spectrum of
the cycloidal trajectories (e.g., Table \ref{table1} and
Fig.~\ref{fig5}).  If higher-order doublets can be seen in the
spectrum (like in Fig.~\ref{fig6}), they can also be identified with
spin waves.  Hence, going e.g.\ from the $3^{\text{rd}}$ to the
$5^{\text{th}}$ level of the hierarchy in the present paper
corresponds to taking into account two additional spin wave modes in
the alternative formalism.  This strong model selection may also be
responsible for inhibiting spin wave emission by the vortex motion as
noted above.

\section*{Acknowledgements}
Work at Los Alamos National Laboratory was sponsored by the United
States Department of Energy.  Travel between Los Alamos and Bayreuth
was partially supported by NATO Collaborative Research Grant No.\
0013/89.


\begin{appendix}
\section*{Appendix}
In this appendix we describe the numerical details of our work.

The time evolution of classical spin systems is governed by the 
Landau-Lifshitz equation
\begin{equation}
\label{LLE}
   \dot{\vec S}_k = \vec S_k\times\vec H_k \quad,\quad
   \vec H_k = -\pdiff{H}{\vec S_k}.
\end{equation}
The index $k$ denotes the site-coordinates of a 2-d square lattice. 
The local field ${\vec H}_k$ is for our model (\ref{eq:hamiltonian}) given by 
\begin{equation}
\label{H_eff}
   {\vec H}_k = J\sum_{(kl)}
       [S_l^x{\vec e}_x + S_l^y{\vec e}_y + (1-\delta)S_l^z{\vec e}_z]
\end{equation}
where here the sum runs over all the nearest neighbors of $k$
not including $k$ itself. 

Equation (\ref{LLE}) with (\ref{H_eff}) was numerically integrated using
a forth-order Runge-Kutta scheme with a time step of 0.04
(in time units $(JS)^{-1}$).
Either free boundary conditions on a circular system, i.e.\ %
$\vec S_{i,j}=0$ for all $(i,j)$ with $i^2 + j^2 > L^2$, or anti-periodic
boundary conditions on a square system were used. 
With $\phi_{i,j}=\tan^{-1}S_{i,j}^y/S_{i,j}^x$ the latter are defined in 
the following way:
\begin{eqnarray*}
   \phi_{i,0} + \phi_{i,L} = \varphi_0 &\:,\:& \phi_{i,1} +
   \phi_{i,L+1} = \varphi_0 \\ \phi_{0,j} + \phi_{L,j} = \varphi_0
   + q_{in}\pi &,& \phi_{1,j} + \phi_{L+1,j} = \varphi_0 +
   q_{in}\pi\\ S^z_{i,0} + S^z_{i,L} = 0 &,& S^z_{i,1} + S^z_{i,L+1} = 0
   \\ S^z_{0,j} + S^z_{L,j} = 0 &,& S^z_{1,j} + S^z_{L+1,j} = 0.
\end{eqnarray*}
$L$ denotes the number of lattice sites in $x$- and $y$-direction,
respectively (i.e.\ the linear dimension of the system).
$q_{in}$ is the total amount of vorticity in the
system ($1$ or $-1$ for one vortex in the system) and 
$\varphi_0$ is an arbitrary constant.

The exact structure of an one-vortex solution is, even in the 
static case, analytically not known. Therefore we calculated 
the initial conditions for our simulations numerically using the
following iteration scheme:
\begin{equation}
\label{D:SSL:IV}    
   \vec{S}^{(n+1)}_k =
       \frac{|\vec{S}_k^{(n)}|}{|\vec{H}_k^{(n)}|}\vec{H}_k^{(n)}.
\end{equation}
Normally this method is used to calculate static solutions
of the Landau-Lifshitz equation. However, iterating only a few
times (a typical value is 30) a pronounced vortex
structure develops which is suitable to serve as the initial
spin field for the simulations.
 
Next we explain how we have determined the vortex position(s)
during a simulation. Using the discrete variant of the contour 
integral $\oint\! d\vec r\: \nabla\phi(\vec r)$
(which yields $2\pi q$ if the contour surrounds the vortex and 
zero otherwise), only the lattice plaquette can be identified
in which the vortex necessarily has to reside.
We estimated the precise position according to the formulas
\begin{mathletters}
\label{X_12}
\begin{eqnarray}
  \textstyle{  
    X_1} &\textstyle{=}&
  \textstyle{
            \frac{1}{2}\frac{-\sin[q^{-1}(\phi_1+\phi_2-\phi_3-\phi_4)]}
                            { \cos[q^{-1}(\phi_1+\phi_2-\phi_3-\phi_4)]
                             -\sin[q^{-1}(\phi_1-\phi_2+\phi_3-\phi_4)]
                             -\cos[q^{-1}(\phi_1-\phi_2+\phi_3-\phi_4)]} }\\
  \textstyle{
    X_2} &\textstyle{=}&  
  \textstyle{
            \frac{1}{2}\frac{ \sin[q^{-1}(\phi_1-\phi_2-\phi_3+\phi_4)]}
                            { \cos[q^{-1}(\phi_1-\phi_2-\phi_3+\phi_4)]
                             +\sin[q^{-1}(\phi_1-\phi_2+\phi_3-\phi_4)]
                             -\cos[q^{-1}(\phi_1-\phi_2+\phi_3-\phi_4)]} }
\end{eqnarray}
\end{mathletters}%
where $\phi_1,\ldots,\phi_4$ label the angles (cf.\ (\ref{motion}a)) 
of the four innermost spins at the vortex core, beginning with the spin
downleft and surrounding the vortex counter-clockwise.
$X_1$ and $X_2$ are measured in units of the lattice constant.
The underlying coordinate system has its origin at the center of the lattice 
plaquette under consideration, i.e.\ $X_1$ and $X_2$ range
between $[-0.5,0.5]$ for `reasonable' values of $\phi_1,\ldots,\phi_4$.

Our basic assumption in deriving (\ref{X_12}) was that
$\phi_1,\ldots,\phi_4$ are distributed according to the static
solution (\ref{eq322}). Considering only differences, this
eliminates the rather annoying constant $\varphi_0$ in (\ref{eq322}),
we can derive the following four equations:
\begin{mathletters}
\label{fourTan}
\begin{eqnarray}
   \tan[q^{-1}(\phi_1-\phi_2)]&=& -\frac{1}{2}\frac{1+2X_2}{X_1^2+X_2^2+X_2} \\ 
   \tan[q^{-1}(\phi_2-\phi_3)]&=& -\frac{1}{2}\frac{1-2X_1}{X_1^2+X_2^2-X_1} \\ 
   \tan[q^{-1}(\phi_3-\phi_4)]&=& -\frac{1}{2}\frac{1-2X_2}{X_1^2+X_2^2-X_2} \\
   \tan[q^{-1}(\phi_4-\phi_1)]&=& -\frac{1}{2}\frac{1+2X_1}{X_1^2+X_2^2+X_1}.
\end{eqnarray}
\end{mathletters}%
We formally assume that the quantities $X_1$, $X_2$ and $X_1^2+X_2^2$
are independent of each other. 
Eqs. (\ref{fourTan}b,d) then constitute a {\em linear} system in $X_1$ and 
$X_1^2+X_2^2$ and analogeously (\ref{fourTan}a,c) in $X_2$ and $X_1^2+X_2^2$. 
Both set of equations can easily be solved and yield after some further
trigonometric steps the above formulas for $X_1$ and $X_2$.  

Of course in general, $\phi_1,\ldots,\phi_4$ are not 
distributed according to  the $\tan^{-1}$-function.
Although then (\ref{fourTan}a-d) do (most probably) not
have a solution at all, the Eqs.\ (\ref{X_12}) can still be justified in the
sense that they provide an approximation of the minimal 
residual of (\ref{fourTan}a-d).
Anyway, the reasonability of (\ref{X_12}) is also acknowledged by
the numerical simulations which show that trajectories are
smoothly connected when the vortex leaves one lattice plaquette
and enters another one. 

\end{appendix}





\clearpage


\begin{figure}
\centerline{\epsfxsize=8.0truecm \epsffile{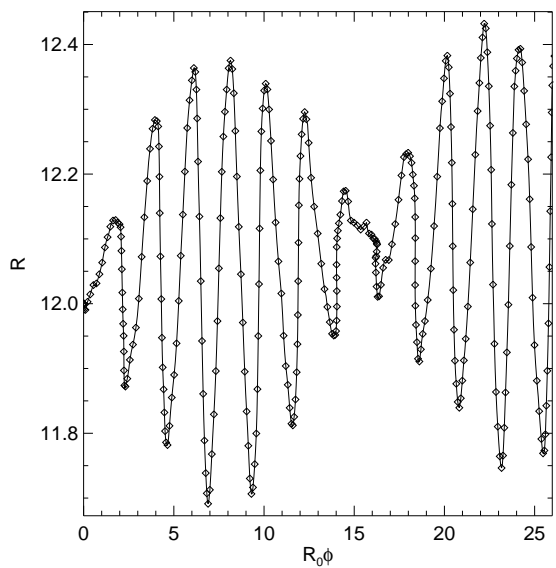}}
\caption{First part of the trajectory of a vortex with $q=p=1$ on a
   circular system with radius $L=36$ and free boundary conditions. 
   The small diamonds ($\diamond$) mark the position
   of the vortex in time intervals of $10(JS)^{-1}$.}
\label{fig4}
\end{figure}

\begin{figure}
\centerline{\epsfxsize=8.0truecm \epsffile{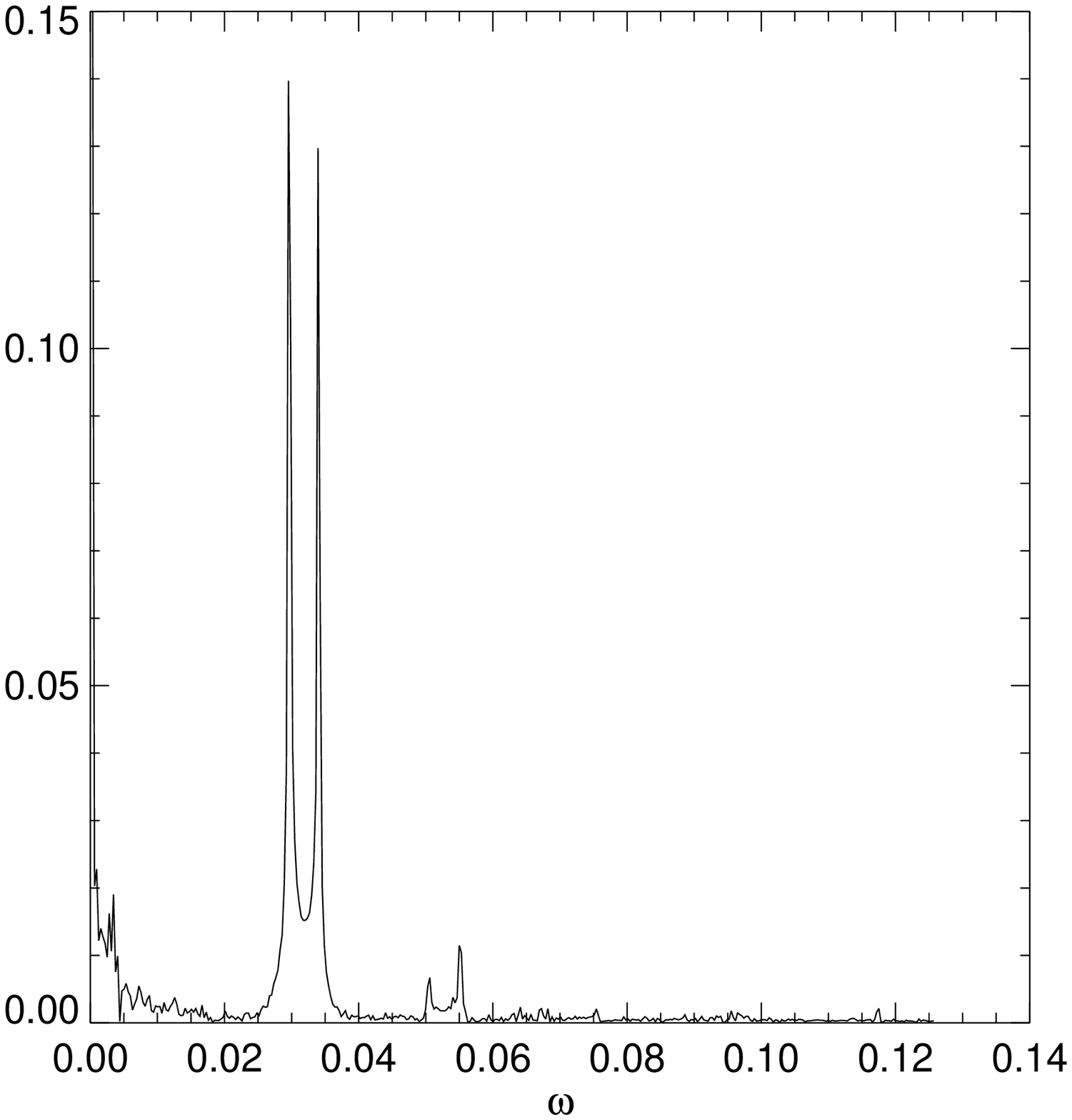} 
            \epsfxsize=8.0truecm \epsffile{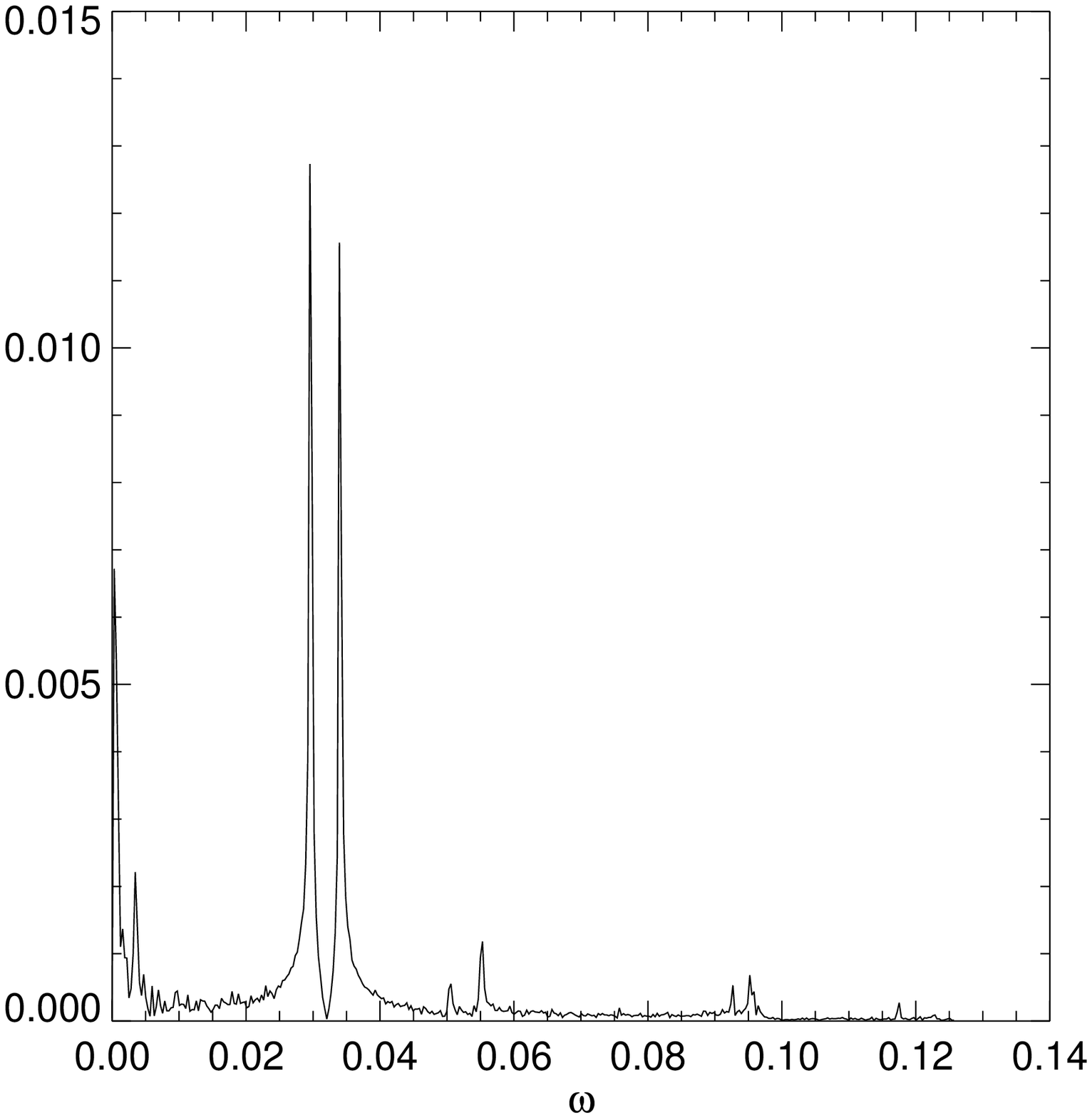}}
\caption{(a) Fourier spectrum of the radial displacements
   $r(t)=R(t)-R_0$, from simulation data for $0\le t\le
   20000(JS)^{-1}$.  The first part of the trajectory is plottet in
   Fig.\ \protect\ref{fig4}.  (b) Spectrum of the aximuthal
   displacements $\varphi(t)=\phi(t)-\omega_0 t$.}
\label{fig5}
\end{figure}

\begin{figure}
\centerline{\epsfxsize=8.0truecm \epsffile{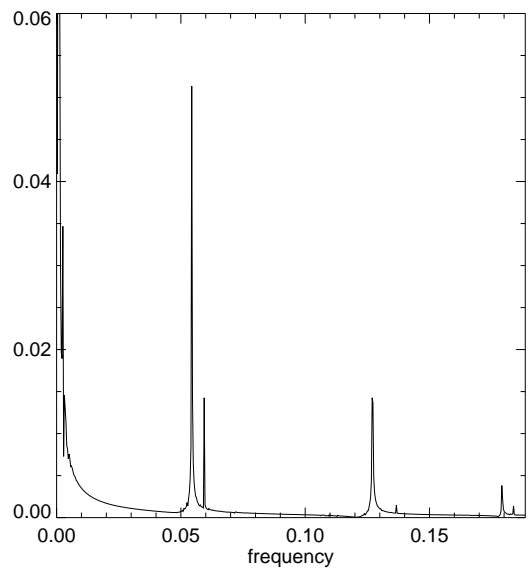}}
\caption{Spectrum of one of the cartesian coordinates of a vortex on a
   $50\times50$ square lattice with antiperiodic boundary conditions,
   from simulation data for $0\le t\le 20000(JS)^{-1}$.}
\label{fig6}
\end{figure}


\clearpage


\begin{table}
\begin{center}
\caption{Observed data from the trajectory of a vortex with $q = p = 1$ 
on a circle of radius $L$.}
\label{table1}
\begin{tabular}{ccccccccc}
$L$ & $R_0$ & $\omega_0/10^{-3}$ & $\omega_1/10^{-2}$ & $\omega_2/10^{-2}$ & 
$\delta_1$ & $\delta_2$ & $\kappa_1$ & $\kappa_2$\\\hline
24 & 4.03 & 1.84 & 4.257 & 5.302 & 1.57 & -1.57 & 1.01 & -1.06\\
24 & 6.02 & 1.90 & 4.257 & 5.276 & 1.56 & -1.56 & 1.03 & -1.11\\
24 & 8.02 & 2.01 & 4.261 & 5.249 & 1.56 & -1.56 & 1.06 & -1.18\\\hline
36 & 5.80 & 0.814 & 2.958 & 3.426 & 1.57 & -1.57 & 1.00 & -1.02\\
36 & 8.28 & 0.823 & 2.962 & 3.417 & 1.52 & -1.50 & 1.02 & -1.09\\
36 & 9.87 & 0.852 & 2.964 & 3.412 & 1.56 & -1.56 & 1.06 & -1.12\\
36 & 12.25 & 0.889 & 2.966 & 3.404 & 1.56 & -1.57 & 1.05 & -1.14\\\hline
72 & 16.11 & 0.205 & 1.546 & 1.661 & 1.43 & -1.42 & -- & -- \\
72 & 24.20 & 0.218 & 1.548 & 1.657 & 1.39 & -1.40 & 1.12 & -1.22\\
72 & 31.95 & 0.243 & 1.550 & 1.653 & 1.45 & -1.46 & 1.24 & -1.37
\end{tabular} 
\end{center}
\end{table}


\begin{table}
\begin{center}
\caption{Parameters $M$ and $A$ of the equation of motion,
   calculated from (\protect\ref{eq561}), (\protect\ref{eq562}). 
   Values in parentheses are extrapolated.}
\label{table2}
\begin{tabular}{ccccc}
   $L$ & $R_0$ & $\eta=R_0/L$ & $M$ & $A$ \\\hline
   24  &  4.03 & 0.168 & 13.7 & 2750 \\
   24  &  6.02 & 0.251 & 12.5 & 2764 \\
   24  &  8.02 & 0.334 & 10.8 & 2776 \\\hline
   36  &  5.80 & 0.161 & 13.9 & 6167 \\
   36  &  8.28 & 0.230 & 12.9 & 6174 \\
   36  &  9.87 & 0.274 & 12.0 & 6180 \\
   36  & 12.25 & 0.340 & 10.7 & 6190 \\\hline
   72  & 12.0  & 0.167 &(13.8)&(24416) \\
   72  & 16.1  & 0.224 & 13.0 & 24436 \\
   72  & 24.2  & 0.336 & 10.7 & 24477 \\
   72  & 31.9  & 0.443 &  7.4 & 24504
\end{tabular}
\end{center}
\end{table}


\begin{table}
\begin{center}
\caption{Parameters $M$ and $A$ of the third-order equation
   of motion, compared to the parameters $M$, $A$, $B$, $C$
   of the fifth-order equation, where $B$ and $C$ belong
   to $\vec X^{(4)}$ and $\vec X^{(5)}$, respectively. 
   The $\omega_\alpha$'s are observed in vortex motion on
   $2L\times2L$ square lattices with antiperiodic
   boundary conditions.}
\label{table3}
\begin{tabular}{ccc}
   $L$ & 10 & 25 \\\hline
   $\omega_1/10^{-1}$ & 1.376 & 0.5435 \\
   $\omega_2/10^{-1}$ & 1.696 & 0.5938 \\
   $\omega_3/10^{-1}$ & 3.749 & 1.2692 \\
   $\omega_4/10^{-1}$ & 4.362 & 1.3670 \\\hline
   $M$ ($3^{\text{rd}}$)       & 8.62  & 9.79 \\
   $A$ ($3^{\text{rd}}$)       & 269   & 1947 \\\hline
   $M$ ($5^{\text{th}}$)       & 10.98 & 13.42 \\
   $A$ ($5^{\text{th}}$)       & 304   & 2300 \\
   $B$ ($5^{\text{th}}$)       & 153   & 1689 \\
   $C$ ($5^{\text{th}}$)       & 1646  & 111991
\end{tabular} 
\end{center}
\end{table}

\end{document}